\documentclass{appolb}
\usepackage{graphicx}
\usepackage{times}
\usepackage[english]{babel}
\usepackage{amsmath}
\usepackage[latin1]{inputenc}
\usepackage[T1]{fontenc}
\usepackage{graphicx}
\usepackage{epsfig}
\usepackage{rotating}
\usepackage{subfigure}
\usepackage{tikz}
\usepackage[subfigure]{ccaption}

\begin{document}
\title{Light quark mass differences in the  $\pi^0 - \eta - \eta'$  system\thanks{Presented at the workshop "Excited QCD 2016", 05-12 March 2016, Costa da Caparica, Portugal. B.H. thanks the organizers for the kind invitation.}%
}
\author{A. A. Osipov, B. Hiller, A. H. Blin, J. Moreira
\address{CFisUC, Department of Physics, University of Coimbra, 3004-516 Coimbra,
Portugal}
}
\maketitle
\begin{abstract}
A generalized  3 flavor Nambu-Jona--Lasinio Lagrangian including the explicit chiral symmetry breaking interactions which contribute at the same order in the large $1/N_c$ counting as the $U_A(1)$  't Hooft flavor determinant is considered to obtain the mixing angles in the  $\pi^0-\eta-\eta'$ system and related current quark mass ratios in close agreement with phenomenological values. 
At the same time an accurate ordering and magnitude of the splitting of states in the low lying pseudoscalar nonet is obtained.
\end{abstract}
\PACS{PACS: 11.30.Rd; 11.30.Qc; 12.39.Fe; 12.40.Yx}

\vspace{0.5cm}
Properties related to the mixing in the $\pi^0-\eta-\eta'$ system are a subject of continued interest, as they address the complexity of non-perturbative QCD subject to the combined effects of chiral symmetry breaking and the $U_A(1)$ anomaly \cite{tHooft:1976}. In addition this system is a standard probe used in  the determination of the values of the current quark masses \cite{Gross:1979},\cite{Feldmann:1999},\cite{Kroll:2005},\cite{Volkov:2012}, and is a process of considerable importance in the studies of weak \cite{Buras:2014} and strong CP violation \cite{Creutz:2004}.

We report on our results for the  $\pi^0-\eta-\eta'$ mixing angles \cite{Osipov:2015} that rely on an effective multi-quark low energy Lagrangian for QCD   \cite{Osipov1:2013}, operational at the scale of spontaneous breaking of chiral symmetry,  of the order of $\Lambda_{\chi SB}\sim 4\pi f_{\pi}$ \cite{Georgi:1984}, and generalizes the Nambu-Jona-Lasinio (NJL) model \cite{Nambu:1961} (where this scale is also related to the gap equation and given by the ultra-violet cutoff $\Lambda$ of the one-loop quark integral) 
as follows: generic vertices $L_i$ of non-derivative type that contribute to the effective potential as $\Lambda\to\infty$\begin{equation}
\label{genL}
   L_i\sim \frac{\bar g_i}{\Lambda^\gamma}\chi^\alpha\Sigma^\beta,
\end{equation}
where powers of $\Lambda$ give the correct dimensionality of the interactions (below we use also unbarred couplings, $g_i=\frac{\bar g_i}{\Lambda^\gamma}$); the $L_i$ are C, P, T and chiral $SU(3)_L\times SU(3)_R$ invariant blocks, built of powers of the  sources $\chi$
which at the end give origin to the explicit symmetry breaking (ESB) and have the same transformation properties as the $U(3)$ Lie-algebra valued field $\Sigma =(s_a-ip_a)
\frac{1}{2}\lambda_a$; here $s_a=\bar q\lambda_aq$, $p_a=\bar q\lambda_ai
\gamma_5q$, and $a=0,1,\ldots ,8$, $\lambda_0=\sqrt{2/3}\times 1$, $\lambda_a$ 
being the standard $SU(3)$ Gell-Mann matrices for $1\leq a \leq 8$.

The interaction Lagrangian without external sources $\chi$ is well known, 
\begin{eqnarray}
\label{L-int}
   L_{int}&=&\frac{\bar G}{\Lambda^2}\mbox{tr}\left(\Sigma^\dagger\Sigma\right)
   +\frac{\bar\kappa}{\Lambda^5}\left(\det\Sigma+\det\Sigma^\dagger\right) 
   \nonumber \\
   &+&\frac{\bar g_1}{\Lambda^8}\left(\mbox{tr}\,\Sigma^\dagger\Sigma\right)^2
   +\frac{\bar g_2}{\Lambda^8}\mbox{tr}
   \left(\Sigma^\dagger\Sigma\Sigma^\dagger\Sigma\right).
\end{eqnarray}  
 The second term is the 't Hooft determinant \cite{Hooft:1976}, 
 the last two the 8 quark ($q$) interactions \cite{Osipov:2006b} which complete the number of relevant vertices in 4D for dynamical chiral symmetry breaking \cite{Andrianov:1993a}.  
The interactions dependent on the sources $\chi$  contain eleven terms \cite{Osipov1:2013} {\footnote{ We omit  $L_1,L_9,L_{10}$ from the list, as they refer to Kaplan-Manohar ambiguity \cite{Manohar:1986} within the model, which allows to set the couplings to 0.}}  
\begin{equation}
  L_\chi =\sum_{i=0}^{10}L_i,
\end{equation}
\begin{eqnarray}
\label{L-chi-1}
   L_0&=&-\mbox{tr}\left(\Sigma^\dagger\chi +\chi^\dagger\Sigma\right), \hspace{0.5cm}
   L_2=\frac{\bar\kappa_2}{\Lambda^3}e_{ijk}e_{mnl}
   \chi_{im}\Sigma_{jn}\Sigma_{kl}+h.c., \hspace{0.5cm}
  \nonumber \\
   L_3&=&\frac{\bar g_3}{\Lambda^6}\mbox{tr}
   \left(\Sigma^\dagger\Sigma\Sigma^\dagger\chi\right)+h.c., \hspace{0.5cm}
   L_4=\frac{\bar g_4}{\Lambda^6}\mbox{tr}\left(\Sigma^\dagger\Sigma\right)
   \mbox{tr}\left(\Sigma^\dagger\chi\right)+h.c., \hspace{0.5cm}
   \nonumber \\
   L_5&=&\frac{\bar g_5}{\Lambda^4}\mbox{tr}\left(\Sigma^\dagger\chi
   \Sigma^\dagger\chi\right)+h.c., \hspace{0.5cm}
   L_6=\frac{\bar g_6}{\Lambda^4}\mbox{tr}\left(\Sigma\Sigma^\dagger\chi
   \chi^\dagger +\Sigma^\dagger\Sigma\chi^\dagger\chi\right), \hspace{0.5cm}
   \nonumber \\
   L_7&=&\frac{\bar g_7}{\Lambda^4}\left(\mbox{tr}\Sigma^\dagger\chi 
   + h.c.\right)^2, \hspace{0.5cm}
   L_8=\frac{\bar g_8}{\Lambda^4}\left(\mbox{tr}\Sigma^\dagger\chi 
   - h.c.\right)^2, \hspace{0.5cm}
\end{eqnarray}

The $N_c$ assignments are
$\Sigma \sim N_c$; $\Lambda \sim N_c^0 \sim 1$; $\chi \sim N_c^0 \sim 1$ \footnote{The counting for $\Lambda$ is a direct consequence of the gap equation $1\sim N_c G\Lambda^2$.}.
We get that exactly the diagrams which survive as $\Lambda\rightarrow\infty$ also surive as $N_c\rightarrow\infty$ and comply with the usual requirements.

At LO in $1/N_c$ only the  $4q$ interactions $(\sim G)$ in eq. (\ref{L-int}) and $L_0$ contribute.
The OZI rule violating vertices are always of order $\frac{1}{N_c}$ with respect to the leading contribution. Non OZI-violating Lagrangian pieces scaling as $N_c^0$ represent NLO contributions with one internal quark loop in 
$N_c$ counting; their couplings encode the admixture of a four quark component ${\bar q}q{\bar q}q$ to the leading ${\bar q}q$ at $N_c\rightarrow\infty$.
Diagrams tracing OZI rule violation are: $\kappa,\kappa_2,g_1,g_4,g_7,g_8$;
Diagrams with admixture of 4 quark and 2 quark states are: $g_2,g_3,g_5,g_6$.

Putting $\chi 
=\frac{1}{2}\mbox{diag}(m_u, m_d, m_s)$,  the current quark masses,
we obtain a consistent set of explicitly breaking chiral symmetry terms. 

 One ends up with 5 parameters needed to describe the LO contributions (the scale $\Lambda$, the coupling $G$, and the $m_i$)  and 10 in NLO ( $\bar\kappa, \bar\kappa_2$, $\bar g_1,\ldots,\bar g_{8}$). 

The details of bosonization in the framework of functional integrals, which lead from  $L=\bar q i\gamma^\mu\partial_\mu q+L_{int}+L_{\chi}$
to the long distance ef\mbox{}fective 
mesonic Lagrangian can be found in \cite{Osipov:2001},\cite{Osipov:2006b},\cite{Osipov1:2013}, here we only collect the result for the kinetic and mesonic pseudoscalar mass terms 

\begin{eqnarray}
\label{mass}
     &&L_{kin}+L_{mass}= \frac{N_c I_1}{8\pi^2}  (\partial \phi_a)^2+\frac{N_cI_0}{4\pi^2}\phi_a^2 \nonumber \\
     &-&\frac{N_cI_1}{24\pi^2}\{[\phi_u^2(2 M_u^2-M_d^2 -M_s^2)+\phi_d^2(2 M_d^2-M_u^2 -M_s^2) \nonumber \\
     &+&\phi_s^2(2 M_s^2-M_u^2 -M_d^2)]\} +\frac{1}{2}\,  h_{ab}^{(2)}\color{black} \phi_a\phi_b +  ...         
\end{eqnarray}
where $h^{(2)}_{ab}$  carries all the dependence on the model couplings and current quark masses and 
$M_i$ $\{i=u,d,s\}$ are the constituent quark masses obtained by solving the model's gap equations. 
The kinetic term requires a redefinition of meson fields
$\phi_a=g \phi_a^R$, $g^2=\frac{4\pi^2}{N_c I_1}=\frac{(M_u+M_d)^2}{2 f_\pi^2}$, which are related to the  flavor $\{u,d,s\}$ and the strange-nonstrange basis as
$ \phi_u=\phi_3+\frac{\sqrt{2}\phi_0 +\phi_8}{\sqrt{3}} 
            =\phi_3+\eta_{ns} $,
$   \phi_d=-\phi_3+\frac{\sqrt{2} \phi_0 +\phi_8}{\sqrt{3}} 
            =-\phi_3+\eta_{ns} $,
$     \phi_s=\sqrt{\frac{2}{3}}\phi_0-\frac{2 \phi_8}{\sqrt{3}}
            =\sqrt{2}\eta_s$.
Defining 
$m_\Delta=\frac{1}{2}(m_d-m_u)$, $m_\Sigma=\frac{1}{2}(m_d+m_u)$, $h_\Delta=\frac{1}{2}(h_d-h_u)$ and $h_\Sigma=\frac{1}{2}(h_d+h_u)$, one has for example for the matrix elements relevant for $\pi^0-\eta$ and $\pi^0-\eta'$ mixing
\begin{eqnarray} 
\label{spameson03}  
&&{\sqrt 6}(h_{03}^{(2)})^{(-1)}= h_\Delta (2 g_2 h_\Sigma + \kappa +  g_3 m_\Sigma)  \nonumber \\
&&+m_\Delta [ g_3 h_\Sigma + 2 (\kappa_2 - g_8 (m_s+2 m_\Sigma) \nonumber \\
&&- (g_5 -g_6) m_\Sigma] 
\end{eqnarray}
and a quite similar expression for  $(h_{38}^{(2)})^{(-1)}$. Note that
both elements vanish in LO in $N_c$ and that explicit  $m_i$ dependence occurs only in presence of ESB interactions at NLO.  In their absence  the effects of ESB are  present in  the  difference of the condensates $h_\Delta\ne 0$ if the conventional
QCD mass term  $m_u\ne m_d$ . In this case only the 't Hooft $\sim \kappa$ and and the 8q $\sim g_2$ contribute to (\ref{spameson03}).

The physical states $\pi_0,\eta,\eta'$ are obtained by diagonalizing the symmetric meson mass matrix  $B_{ab}$ in the subspace $\{0,3,8\}$ of (\ref{mass}). Since one is free to transform between the states, we chose to represent the mixing angles with the strange-nonstrange basis as reference{\footnote {We show that the decay constants transform as the states in this basis in our model \cite{Osipov:2015}.}
 by the following transformation   ${\cal S}={\cal U}{\cal V}$
\begin{eqnarray}
\label{trans}
&& \left(\phi_3,\phi_0,\phi_8\right)S^{-1} S  \left(\begin{array}{ccc} B_{33} & B_{03} & B_{38} \\
                            B_{03} & B_{00} & B_{08} \\
                            B_{38} & B_{08} & B_{88} \\  
   \end{array} \right)S^{-1} S \left(\begin{array}{c} \phi_3 \\ \phi_0 \\ \phi_8 \\
         \end{array} \right), \nonumber 
\end{eqnarray} 
 rotating first to this basis through the orthogonal involutory matrix ${\cal V}$
\begin{eqnarray}
\label{sns}
&&\left(\begin{array}{c} \phi_3 \\ \eta_{ns} \\ \eta_s
         \end{array} \right)= {\cal V} \left(\begin{array}{c} \phi_3  \\ \phi_0 \\ \phi_8
         \end{array} \right);  
 \qquad{\cal V}= \frac{1}{\sqrt{3}}\left(\begin{array}{ccc} \sqrt{3} & 0 & 0 \\
                            0 & \sqrt{2}  & 1 \\
                            0 & 1 & -\sqrt{2} \nonumber \\  
  \end{array} \right)
\end{eqnarray}
and then using the unitary transformation ${\cal U}$ to obtain the physical states \cite{Kroll:2005}
\begin{equation}
\label{lin}
    \left( \begin{array}{c} \pi^0 \\
                            \eta  \\
                            \eta'
           \end{array} \right)
    = {\cal U}(\epsilon_1,\epsilon_2,\psi )
\left( \begin{array}{c} \phi_3 \\
                        \eta_{ns}  \\
                        \eta_{s}
           \end{array} \right),
\end{equation}
${\cal U}$ is linearized in the $\pi^0-\eta$ and $\pi^0-
\eta'$ mixing angles, since it is assumed that $\phi_3$ couples weakly to the $\eta_{ns}$ and $\eta_s$
states, decoupling in the isospin limit, while the mixing for the $\eta-\eta'$ system is
strong {\footnote {The usual redefinitions $\epsilon =\epsilon_2 +\epsilon_1\cos\psi, \epsilon'=\epsilon_1\sin\psi$ are adopted.}} ,
\begin{equation}
\label{Um}
   {\cal U} = \left(\begin{array}{ccc} 1& \epsilon_1 +\epsilon_2\cos\psi
                                & -\epsilon_2\sin\psi \\
            -\epsilon_2-\epsilon_1\cos\psi& \cos\psi& -\sin\psi \\
            -\epsilon_1\sin\psi & \sin\psi& \cos\psi
            \end{array} \right).
\end{equation}
%
%
%
We checked this hypothesis in our model calculations by diagonalizing the mass matrix also exactly using the explicit analytical expressions for the \emph{eigenvalues} of a symmetric $3\times3$ matrix $ M$ \cite{Smith}
\begin{align}
\lambda_1=&\xi-\sqrt{\varsigma} \left(\cos\left[\varphi\right]+\sqrt{3}\sin\left[\varphi\right]\right)\nonumber\\
\lambda_2=&\xi-\sqrt{\varsigma} \left(\cos\left[\varphi\right]-\sqrt{3}\sin\left[\varphi\right]\right)\nonumber\\
\lambda_3=&\xi+2 \sqrt{\varsigma}\cos\left[\varphi\right],\nonumber
\end{align}
where we use the abbreviations:
$\xi =\frac{\mathrm{tr}\left[M\right]}{3}, \mathcal{M} =M-\xi \mbox {\bf I},$ 
$\varsigma  =\frac{1}{6}\sum_i\sum_j \left(\mathcal{M}_{ij}\right)^2,$
$ \vartheta  =\frac{1}{2}\mathrm{det}\left[\mathcal{M}\right]\nonumber,
 \varphi    =\frac{1}{6}\mathrm{ArcTan}\left[\frac{\sqrt{\varsigma^3-\vartheta}}{\vartheta}\right].
$
The \emph{eigenvectors} can then be obtained by normalizing the vectors given by:
$\vec{v_i}=\left(\left(\overrightarrow{M_{1}}-\lambda_i \hat{e}_1\right)\times\left(\overrightarrow{M_{2}}-\lambda_i\hat{e}_2\right)\right)^\ast$, 
where $\overrightarrow{M_{j}}$ corresponds to the $j$ column of $M$. 
These are then used to build the diagonalization matrix with the standard parametrization of the CKM matrix (with the abbreviation $C_{ij}\equiv \cos\left[\theta_{ij}\right]$, $S_{ij}\equiv \sin\left[\theta_{ij}\right]$,
$\theta_{12}=-\epsilon$,  $\theta_{13}=-\epsilon'$, $\theta_{23}=\psi$ and phase=0)

{
\begin{align}
U=
\left[\begin{array}{c c c}
 C_{12}C_{13}                    & -S_{12}C_{23}-C_{12}S_{13}S_{23}  &  S_{12}S_{23}-C_{12}S_{13}C_{23}       \\
 S_{12}C_{13}                    &  C_{12}C_{23}-S_{12}S_{13}S_{23}  & -C_{12}S_{23}-S_{12}S_{13}C_{23}\\
 S_{13}                          &  C_{13}S_{23}                     &  C_{13}C_{23}
\end{array}\right].\nonumber
\end{align}
}
\noindent The numerical deviations, as compared to (\ref{Um}) are within $2\%$ for the cases considered.
The results are shown in the Tables 1-3. One sees that
 the explicit symmetry breaking interactions of the generalized NJL Lagrangian considered are crucial to obtain the phenomenological quoted value for the ratio $\frac{\epsilon}{\epsilon'}$, Table 4.
We obtain values for the $\epsilon$ mixing angle which lie within the results discussed in the literature.
Unfortunately the value for $\epsilon'$ is much less discussed.
The values $\epsilon$ and $\epsilon'$ are  reasonably close to the ones indicated in \cite{Feldmann:1999}, \cite{Kroll:2005} for the renormalization group invariant mass ratio $m_u/m_d$ and current quark mass values in  agreement with the presently quoted average values, 
$\frac{m_u}{m_d}=0.46(5)$,  $m_u=2.15(15) MeV$ , $m_d=4.70(20) MeV$, $m_s=93.5\pm 2.5 MeV$ \cite{PDT:2014}.

\begin{table*}
\small
\caption{Empirical fits (input marked by (*), also in Tables \ref{ibpar},\ref{isobreak}) for set A (with NLO ESB interactions), and  B (LO ESB). Masses in units of MeV, angle $\psi$ in degrees. 
}
\label{mi}
\begin{tabular*}{\textwidth}{@{\extracolsep{\fill}}lrrrrrrrrl@{}}
\hline
Set & \multicolumn{1}{c}{${m_\pi^0}$}
     & \multicolumn{1}{c}{${m_\pi^{\pm}}$}
     & \multicolumn{1}{c}{$m_\eta$}
     & \multicolumn{1}{c}{${m_\eta'}$}
     & \multicolumn{1}{c}{$m_K^0$}
     & \multicolumn{1}{c}{$m_K^{\pm}$}
     & \multicolumn{1}{c}{$f_\pi$}
     & \multicolumn{1}{c}{$f_K$}
     & \multicolumn{1}{c}{$\psi$} \\
\hline
A  & 136* & 136.6 & 547* & 958* & 500 & 494* & 92*  & 113*  & 39.7*  \\
B    & 136* & 137.0 & 477 & 958* & 501& 497*  & 92*  & 116*   & 39.7*   \\
\hline
\end{tabular*}
\end{table*}
\begin{table*}
\small
\caption{The couplings emerging from the fits have the following units: $[G]=$ GeV$^{-2}$,
         $[\kappa ]=$ GeV$^{-5}$, $[g_1]=[g_2]=$ GeV$^{-8}$, $[\kappa_2]=$ GeV$^{-3}$,
$[g_3]=[g_4]=$ GeV$^{-6}$, $[g_5]=[g_6]=[g_7]=[g_8]=$ GeV$^{-4}$. $\Lambda=828.5^*; 835.7$ MeV in A,B. 
     }
\label{ibpar}
\begin{tabular*}{\textwidth}{@{\extracolsep{\fill}}lrrrrrrrrrrl@{}}
\hline
Set & \multicolumn{1}{c}{${G}$}
     & \multicolumn{1}{c}{$-\kappa$}
     & \multicolumn{1}{c}{${g_1}$}
     & \multicolumn{1}{c}{$g_2$}
     & \multicolumn{1}{c}{$\kappa_2$}
     & \multicolumn{1}{c}{$g_3$}
     & \multicolumn{1}{c}{$g_4$}
     & \multicolumn{1}{c}{$g_5$} 
     & \multicolumn{1}{c}{$g_6$}
     & \multicolumn{1}{c}{$g_7$}
     & \multicolumn{1}{c}{$g_8$} \\
\hline
A  & 10.48 & 116.8 & 3284 & 1237   & 6.24  & 2365   & 1182  &160 & 712 & 580 & 44  \\
B  & 9.79 & 137.4 & 2500* &   117    & 0*   & 0*  &0* &0* & 0*& 0* & 0* \\
\hline
\end{tabular*}
\end{table*}



\begin{table}[!ht]
\small
\caption{ $\frac{m_u}{m_d}=0.46^*$, current and constituent quark masses  $m_u$, $m_d$, $m_s$, $M_u$, $M_d$, $M_s$  in MeV,   $\pi^0-\eta$, $\pi^0-\eta'$ mixing angles  $\epsilon$ and $\epsilon'$. }
\label{isobreak}
\begin{tabular*}{\textwidth}{@{\extracolsep{\fill}}lrrrrrrrrrl@{}}%
\hline
Set 
      & \multicolumn{1}{c}{$ m_u$}
      & \multicolumn{1}{c}{$m_d$}
      & \multicolumn{1}{c}{$m_s$}
      & \multicolumn{1}{c}{$M_u$}
      & \multicolumn{1}{c}{$M_d$}
      & \multicolumn{1}{c}{$M_s$}
      & \multicolumn{1}{c}{$\epsilon$}
      & \multicolumn{1}{c}{$\epsilon'$} 
      & \multicolumn{1}{c}{$\frac{\epsilon}{\epsilon'}$} \\
\hline
A  
 & 2.179   & 4.760 & 95*& 372  & 375 &544*& 0.014*  &  0.0037*  & 3.78   \\
B  
& 3.774   & 8.246 & 194 & 373  & 380 &573  & 0.022  & 0.0025  & 8.78   \\
\hline
\end{tabular*}
\end{table}

\begin{table}[!ht]
\small
\caption{ $\epsilon$ and $\epsilon'$ values in the literature. }
\label{epp}
\begin{tabular*}{\textwidth}{@{\extracolsep{\fill}}lrrrrrrrrl@{}}
\hline
      & \multicolumn{1}{c}{$\epsilon$} 
      & \multicolumn{1}{c}{$\epsilon'$}
      & \multicolumn{1}{c}{$\frac{\epsilon}{\epsilon'}$} \\
\hline
\cite{Feldmann:1999} phen. \color{black}   & 0.014 & 0.0037 & 3.78     \\
\cite{Kroll:2005} phen. \color{black} & $0.017\pm 0.002$ & $0.004\pm 0.001$ & $4.25\pm 1.17$    \\
\cite{Goity:2002} ChPt NLO  \color{black}& $0.014\div 0.016$ & - & -    \\  
\cite{Coon:1986} phen.  \color{black} & 0.021 & - & -   \\  
\cite{BES:2004} Exp.  \color{black} & $0.030 \pm 0.002$ & - &- \\  
\cite{Tippens:2001} Exp.  \color{black}  & $0.026\pm 0.007$ & - & -   \\  
\hline
\end{tabular*}
\end{table}


\begin{thebibliography}{99}
\bibitem{tHooft:1976} G. 't Hooft, Phys. Rev. Let.  \textbf{37}, 8 (1976); G. 't Hooft, Phys. Rept. 142, 357 (1986); G. Veneziano, Nucl. Phys. B  \textbf{159}, 213 (1979), E. Witten, Annals of Phys. 128, 863 (1980).
\bibitem{Gross:1979} D. J. Gross, S. B. Treiman, F. Wilczek,  Phys. Rev. D \textbf{19}, 2188 (1979); J. Gasser, H. Leutwyler, Annals of Phys. \textbf{158}, 142 (1984); J. Gasser, H. Leutwyler,  Nucl. Phys. B \textbf{250}, 465 (1985).
\bibitem{Feldmann:1999} T. Feldmann, P. Kroll, B. Stech, Phys. Lett. B \textbf{449}, 339 (1999).
\bibitem{Kroll:2005} P. Kroll, Int. J. Mod. Phys. A {\bf 20}, 331 (2005), ibid. Mod.Phys.Lett. A20,  2667 (2005)
\bibitem{Volkov:2012} M.K. Volkov, D.G. Kostunin, Phys. Rev. D, {\bf 86},  013005 (2012)
\bibitem{Buras:2014} A. J. Buras, J.-M. G\'erard, W. A. Bardeen, Europ. Phys. J C \textbf{74}, 2871 (2014).
\bibitem{Creutz:2004} R. F. Dashen, Phys. Rev. D \textbf{3} , 1879 (1971); M. Creutz, Phys. Rev. Let. \textbf{92}, 201601 (2004).
\bibitem{Osipov:2015} A.A. Osipov, B. Hiller, . H. Blin, arxiv:1501.06146 [hep-ph].
\bibitem{Osipov1:2013} A. A. Osipov, B. Hiller, A. H. Blin, Eur. Phys. J. A
   {\bf 49}, 14 (2013);  A. A. Osipov, B. Hiller, A. H. Blin,  Phys. Rev. D   {\bf 88}, 054032 (2013).
\bibitem{Georgi:1984} A. Manohar, H. Georgi, Nucl. Phys. B \textbf{234}, 189
   (1984).
\bibitem{Nambu:1961} Y. Nambu, G. Jona-Lasinio,
   Phys. Rev. \textbf{122}, 345 (1961); {\it ibid.} \textbf{124}, 246 (1961).
\bibitem{Hooft:1976} G. 't Hooft, Phys. Rev. D \textbf{14}, 3432 (1976); {\it ibid.}  \textbf{18}, 2199 (1978).
\bibitem{Osipov:2006b} A. A. Osipov, B. Hiller, J. da Provid\^encia, Phys.
   Lett. B \textbf{634}, 48 (2006); A. A. Osipov, B. Hiller, A. H. Blin,
   J. da Provid\^encia, Annals of Phys. \textbf{322}, 2021 (2007). 
\bibitem{Andrianov:1993a} A. A. Andrianov, V. A. Andrianov, Theor. Math.
   Phys. \textbf{94}, 3 (1993);  A. A. Andrianov, V. A. Andrianov,  Int. J. of Mod.
   Phys. A \textbf{8}, 1981 (1993).
\bibitem{Manohar:1986} D. B. Kaplan, A. V. Manohar, Phys. Rev. Lett. 
   \textbf{56}, 2004 (1986).
\bibitem{Osipov:2001} A. A. Osipov, B. Hiller, Phys. Lett. B \textbf{515}, 458; A. A. Osipov, B. Hiller, Phys. Rev. D \textbf{64} 087701 (2001); A. A. Osipov, B. Hiller,  Europ. Phys. J.  C   \textbf{35},  223 (2004); A. A. Osipov, B. Hiller. J. Moreira, A. H. Blin,  Europ. Phys. J.  C   \textbf{46},  225 (2006); 
\bibitem{Smith} O. K. Smith, Commun. ACM 4, 168 (1961), ISSN 0001-0782;
Joachim Kopp , Int.J.Mod.Phys.C19:523-548,2008
\bibitem{Goity:2002} J. L. Goity, A. M. Bernstein, B. R. Holstein, Phys Rev. D \textbf{66}, 076014
   (2002).
\bibitem{Coon:1986} S.A. Coon, B.H.J. Mc Kellar, M. D. Scadron, Phys.Rev. D \textbf{34}, 2784 (1986).
\bibitem{BES:2004} J. Z. Bai {\it et. al.} [BES Collaboration], Phys. Rev. D \textbf{70} 012006 (2004).
\bibitem{Tippens:2001} W. B. Tippens {\it et. al.}, Phys. rev. D \textbf{63} 052001 (2001).
\bibitem{PDT:2014} K.A. Olive et al., Particle Data Group, Chinese Phys. C {\bf 38}, 090001 (2014).
\end{thebibliography}
\end{document}